\documentclass[preprint,3p]{elsarticle}
\usepackage[fleqn]{amsmath}
\usepackage{amssymb,amsthm}
\usepackage{mathtools}
\usepackage{mathrsfs}
\usepackage{bm}
\usepackage{slashed}	
\usepackage{soul}
\usepackage{hyperref}
\hypersetup{colorlinks=true}
\usepackage{lineno}
\biboptions{sort&compress}
\usepackage{verbatim}
\usepackage{fullpage}
\usepackage{graphicx}
\usepackage{tikz}
\usepackage{subfig} 	
\usepackage{relsize}	
\usepackage{float}
\usepackage{color}
\usepackage{xcolor}
\usepackage{array}
\usepackage{booktabs}
\usepackage{arydshln}
\usepackage{lscape}
\usepackage{multirow}

\def\dd{{\mathrm{d}}}

\mathchardef\-="2D

\makeatletter

\newcommand{\Rmnum}[1]{\expandafter\@slowromancap\romannumeral #1@}
\newcommand{\dif}{\mathrm{d}}
\makeatother

\colorlet{darkgreen}{green!50!black}
\colorlet{brightyellow}{yellow!75!red}
\colorlet{orange}{red!50!yellow}
\colorlet{darkblue}{blue!60!black}
\colorlet{darkred}{red!80!black}

\journal{}

\begin{document}
\begin{frontmatter}

\title{Diffractive charmonium spectrum in high energy collisions in the basis light-front quantization approach}
\author{Guangyao~Chen\corref{ca}}
\ead{gchen@iastate.edu}

\author{Yang~Li\corref{ca2}}
\ead{leeyoung@iastate.edu}

\author{Pieter~Maris\corref{ca3}}
\ead{pmaris@iastate.edu}

\author{Kirill~Tuchin\corref{ca4}}
\ead{tuchin@iastate.edu}

\author{James~P.~Vary\corref{ca5}}
\ead{jvary@iastate.edu}

\cortext[ca2]{Corresponding author}

\address{Department of Physics and Astronomy, Iowa State University, Ames, IA 50011, USA}

\begin{abstract}

Using the charmonium light-front wavefunctions obtained by diagonalizing an effective Hamiltonian with the one-gluon exchange interaction 
and a confining potential inspired by light-front holography in the basis light-front quantization formalism, we compute production of
charmonium states in diffractive deep inelastic scattering and ultra-peripheral heavy ion collisions within the dipole picture. Our method allows us to predict yields of all vector charmonium states below the open flavor thresholds in high-energy deep inelastic scattering, proton-nucleus and ultra-peripheral heavy ion collisions, without introducing any new parameters in the light-front wavefunctions. The obtained charmonium cross section is in reasonable agreement with
experimental data at HERA, RHIC and LHC. We observe that the cross-section ratio $\sigma_{\Psi(2s)}/\sigma_{J/\Psi}$ reveals significant
independence of model parameters.   

\end{abstract}

\end{frontmatter}

\section{Introduction} \label{sec:intro}

Exclusive vector meson production in diffractive deep inelastic scattering (DIS) and deeply virtual Compton scattering (DVCS) are effective tools for 
studying Quantum Chromodynamics (QCD) \cite{Gribov:1984tu}. At low $x$ these processes are dominated by gluon saturation
\cite{JalilianMarian:1996xn,Gelis:2010nm}. Models incorporating saturation physics successfully describe the high precision data
harvested at the Hadron-Electron Ring Accelerator (HERA)
\cite{Golec-Biernat:1998js,Levin:2000mv,Gotsman:2001ne,Kowalski:2003hm,Gotsman:2003ww,Iancu:2003ge} and are instrumental for deriving
predictions for future experiments at the Large Hadron electron Collider (LHeC) \cite{AbelleiraFernandez:2012cc} and the Electron-Ion Collider (EIC) \cite{Accardi:2012qut}. Theoretical calculations often
employ the dipole model \cite{Mueller:1989st,Nikolaev:1990ja} that relies on the separation of scales: in the proton rest frame, the
lifetime of the virtual photon and the quarkonium formation time are much longer than the time scale of the interaction.
The dipole model was used in Refs.~\citep{Kowalski:2006hc, Marquet:2007nf} to describe  both exclusive and diffractive HERA measurements at low $x$.

The largest theoretical uncertainty in the calculation of diffractive heavy quarkonium production in the dipole picture arises from poor 
knowledge of the heavy quarkonium light-front wavefunction (LFWF). In phenomenological applications, the LFWFs are simply educated guesses
with several free parameters \citep{Kowalski:2003hm,Kowalski:2006hc}. While such phenomenological models can be successful in explaining the
experimental data, the presence of free parameters limits the predictive power. With electron-ion colliders on the horizon, where around
$30\%$ of total events are expected to be diffractive, finding well-constrained heavy quarkonium LFWFs based on the dynamics of QCD
becomes important.      

Fortunately, recent progress in the basis light-front quantization (BLFQ) approach \cite{Honkanen:2010rc, Vary:2009gt,Zhao:2014xaa, Wiecki:2014ola, Adhikari:2016idg} has 
paved an avenue for improving the understanding of the heavy quarkonium system. It has enabled the computation of the LFWFs for any heavy
quarkonium state and thus calculate the corresponding diffractive cross sections. The BLFQ approach has been successfully applied to
calculate the electron anomalous magnetic moment \cite{Zhao:2014xaa}, and to study the positronium system
\cite{Wiecki:2014ola,Adhikari:2016idg}. Recently, some of us employed the light-front Hamiltonian formalism to obtain the mass spectra and LFWFs for
charmonium and bottomonium \cite{Li:2015zda}. This was achieved by diagonalizing an effective Hamiltonian that incorporates the one-gluon exchange interaction
and a confining potential inspired by light-front holography \cite{deTeramond:2005su,Brodsky:2014yha}. The decay constants and the charge form factors for selected
eigenstates calculated using these LFWFs are comparable to the experimental measurements as well as to results from Lattice QCD
and Dyson-Schwinger Equation approaches. Compared to phenomenological LFWFs used in the literature, LFWFs from the BLFQ approach possess
appealing merits. In particular, the BLFQ LFWFs arise from successful fits to the heavy quarkonia mass spectroscopy, show success in applications
to decay constants and provide predictions for additional quantities such as charge form factors all within the same formalism. 

The main goal of this letter is to employ the
theoretically sound and phenomenologically-constrained BLFQ wavefunctions to compute the diffractive cross sections for the heavy quarkonium production at low $x$ using the dipole 
model to take into account the gluon saturation.

\section{Background} 
\label{sec:background}
In the dipole model, the amplitude for  exclusive heavy quarkonium production  in DIS can be calculated as \cite{Kowalski:2006hc}
\begin{eqnarray}
  \mathcal{A}^{\gamma^* p\rightarrow Ep}_{T,L}(x,Q,\Delta) = \mathrm{i}\,\int\!\dif^2\bm{r}\int_0^1\!\frac{\dif{z}}{4\pi}\int\!\dif^2\bm{b}
\;(\Psi_{E}^{*}\Psi)_{T,L} (r,z,Q) \; 
  \mathrm{e}^{-\mathrm{i}[\bm{b}-(1-z)\bm{r}]\cdot\bm{\Delta}}
  \;\frac {\dif\sigma_{q\bar q}}{\dif^2 \bm b} (x,r) \; ,
  \label{eq:newampvecm}
\end{eqnarray}
where $T$ and $L$ denote the transverse and longitudinal polarization of the virtual photon (with virtuality $Q^2$) and the produced 
quarkonium, and $t= - \bm{\Delta}^2$ denotes the momentum transfer. On the right-hand side, $\bm{r}$ is the transverse size of the color
dipole, $z$ is the LF longitudinal momentum fraction of the quark, $\bm{b}$ is the impact parameter of the dipole relative to the proton and
$x$ is the Bjorken variable. $\Psi$ and $\Psi_{E}^{*}$ are LFWFs of the virtual photon and the exclusively produced quarkonium respectively.
The cross section is related to the amplitude via 
\begin{eqnarray}
\frac{\dif \sigma^{\gamma^* p\rightarrow Ep}_{T,L}}{\dif t} = \frac{1}{16 \pi} \vert \mathcal{A}^{\gamma^* p\rightarrow
Ep}_{T,L}(x,Q,\Delta)  \vert^2 \; .
\end{eqnarray}
Furthermore, several phenomenological corrections are needed in order to describe the experimental data. For example, the contribution from
the real part of the scattering amplitude is conventionally incorporated by multiplying the cross section by a factor
$(1+\beta^2)$~\cite{Kowalski:2006hc}, where $\beta$ is the ratio of the real and imaginary parts of the scattering amplitude, and is calculated as \cite{Ryskin:1995hz}
\begin{eqnarray}
  \beta = \tan(\pi\lambda/2), \quad\text{with}\quad \lambda \equiv \frac{\partial\ln\left(\mathcal{A}_{T,L}^{\gamma^* p\rightarrow Ep}
\right)}{\partial\ln(1/x)}.
  \label{eq:beta}
\end{eqnarray}
The skewedness correction, which takes into account the fact that two gluons interacting with the dipole are carrying slightly different 
momentum fractions, will be specified in Sec.~\ref{ssec:dipole}, since it has been implemented differently for different dipole models in
the literature.  

\begin{table}
  \centering
  \begin{tabular}{cccccccccc}
    \hline\hline
    Model &  $Q^2$/GeV$^2$ & $N_f$ & $m_{u,d,s}$/GeV & $m_c$/GeV & $\mu_0^2/\mathrm{GeV}^2$ & $A_g$ & $\lambda_g$ & $\chi^2/\text{d.o.f.}$ 
\\ \hline
    bSat I  &  [0.25,650] & 3 &$0.14$ & $1.4$ & $1.17$ & $2.55$ & $0.020$ & $193.0/160=1.21$ \\
    bSat II  &  [0.25,650] & 3 & $0.14$ & $1.35$ & $1.20$ & $2.51$ & $0.024$ & $190.2/160=1.19$ \\
    bSat III  &  [0.25,650] & 3 & $0.14$ & $1.5$ & $1.11$ & $2.64$ & $0.011$ & $198.1/160=1.24$ \\
    bSat IV  &  [0.75,650] & 4 & $\approx 0$ & $1.27$ & $1.51$ & $2.308$ & $0.058$ & $298.89/259=1.15$ \\
    bSat V  &  [0.75,650] & 4 & $\approx 0$ & $1.4$ & $1.11$ & $2.373$ & $0.052$ & $316.61/259=1.22$ \\
 \hline\hline
  \end{tabular}
  \caption{Parameters of the initial gluon distribution in the bSat model in Eq.~\eqref{eq:inputgluon} determined from fits to $F_2$ data. 
Parameters of the bSat I-III are fitted to ZEUS data only \cite{Kowalski:2006hc}. Parameters of the bSat IV and V are fitted to combined
HERA data \cite{Rezaeian:2012ji}
.}
  \label{tab:bsat}
\end{table}

\begin{table}
  \centering
  \begin{tabular}{ccccccccc}
    \hline\hline
    Model &  $B_\text{CGC}$/GeV$^{-2}$ & $m_c$/GeV & $\gamma_s$ & $\mathcal{N}_0$ & $x_0$ & $\lambda_s$ & $\chi^2/\text{d.o.f.}$ \\ \hline
    bCGC I  &  $5.591$ & $1.4$ & $0.7376$ & $0.7$ & $1.632\times10^{-5}$ & $0.2197$ & $144.0/160=0.900$ \\
    bCGC II  &  $5.5$ & $1.27$ & $0.6599$ & $0.3358$ & $0.00105$ & $0.2063$ & $368.4/297=1.241$ \\
    bCGC III  & $5.5$ & $1.4$ & $0.6492$ & $0.3658$ & $0.00069$ & $0.2023$ & $370.9/297=1.249$ \\
 \hline\hline
  \end{tabular}
  \caption{Parameters of the bCGC model in Eq.~\eqref{eq:bcgc} determined from fits to $F_2$ data. Parameters in the bCGC I are fitted to 
ZEUS data only \cite{Soyez:2007kg}. Parameters of the bCGC II and III are fitted to combined HERA data \cite{Rezaeian:2013tka}
.}
  \label{tab:bCGC}
\end{table}

\subsection{Dipole cross section parametrizations} 
\label{ssec:dipole}

There are many dipole cross section parametrizations available in the literature based on different theoretical considerations and inspired 
by the Golec-Biernat Wuesthoff (GBW) model \cite{Golec-Biernat:1998js}. For this study we employ two representative dipole parametrizations: the impact parameter
dependent saturation model (bSat) \cite{Kowalski:2003hm} and the impact parameter dependent Color Glass Condensate model (bCGC)
\cite{Iancu:2003ge} to take advantage of their explicit impact parameter dependence, which is important in diffractive quarkonium
production.

The bSat dipole model is based on the Glauber-Mueller formula \cite{Mueller:1989st} and assumes the dipole cross section as follows,
\begin{eqnarray}
\label{eq:bsat}
  \frac{\dif\sigma_{q\bar{q}}}{\dif^2\bm{b}} = 2\left[1-\exp\left(-\frac{\pi^2}{2N_c}r^2\alpha_s(\mu^2)xg(x,\mu^2)T(b)\right)\right],
\end{eqnarray}
where $T(b)$ is the proton shape function, which is assumed to be Gaussian, $T_G(b) =\exp(-b^2/2B_G)/(2\pi B_G)$, with $B_G = 4$~GeV$^{-2}$. 
$\alpha_s$ is determined using LO evolution of the running coupling, with fixed number of flavors $N_f$. $\mu^2$ is related to the
dipole size $r$ through $\mu^2=4/r^2+\mu_0^2$. The gluon density is determined using LO Dokshitzer-Gribov-Lipatov-Altarelli-Parisi evolution \cite{Bartels:2002cj} from an
initial scale $\mu_0^2$, where the initial gluon density is,
\begin{eqnarray}
 \label{eq:inputgluon}
  xg(x,\mu_0^2) = A_g\,x^{-\lambda_g}\,(1-x)^{5.6}. 
 \end{eqnarray} 
In the bSat dipole model, $\mu_0$, $A_g$ and $\lambda_g$ are parameters to be determined by the inclusive DIS data \cite{Chekanov:2001qu,Abramowicz:1900rp,Abramowicz:2015mha}. We use parametrizations
given in Refs.~\cite{Kowalski:2006hc,Rezaeian:2012ji} for this investigation, which we provide in Table~\ref{tab:bsat}. We follow the
prescription in Ref.~\cite{Kowalski:2006hc} for the skewedness correction in the bSat dipole model. $R_\text{bSat}$ is assumed to be
\begin{eqnarray}
  R_\text{bSat}(\delta_{bSat}) = \frac{2^{2\delta_\text{bSat}+3}}{\sqrt{\pi}}
\frac{\Gamma(\delta_\text{bSat}+5/2)}{\Gamma(\delta_\text{bSat}+4)} \quad\text{with}\quad \delta_\text{bSat}  \equiv
\frac{\partial\ln\left[xg(x,\mu^2)\right]}{\partial\ln(1/x)}.
 \label{eq:Rg_bsat}
\end{eqnarray}
The obtained $R_\text{bSat}$ is then applied multiplicatively to the gluon density function in Eq.~(\ref{eq:bsat}). This prescription of the skewedness correction is also adopted in Refs.~\cite{Lappi:2010dd,Toll:2012mb,Lappi:2013am,Rezaeian:2012ji}.

The bCGC dipole model is a smooth interpolation of the solutions of the Balitsky-Fadin-Kuraev-Lipatov equation \cite{BFKL} for small dipole sizes and the 
Levin-Tuchin solution \cite{Levin:1999mw} of the Balitsky-Kovchegov equation \cite{BK} deep inside the saturation region for larger dipoles,  
\begin{eqnarray}
\label{eq:bcgc}
\frac {\dif\sigma_{q\bar q}}{\dif^2\bm{b}}=2 \mathcal{N}(r Q_s,x) =2 
  \begin{cases}
    \mathcal{N}_0 \left(\frac{rQ_s}{2}\right)^{2(\gamma_s + \frac{1}{\kappa_s \lambda_s \ln (1/x)} \ln \frac{2}{rQ_s})}  & :\quad rQ_s\le 2\\
    1-\mathrm{e}^{-\mathcal{A} \ln^2(\mathcal{B} rQ_s)} & :\quad rQ_s>2
  \end{cases},
\end{eqnarray}
with $Q_s\equiv Q_s(x)=(x_0/x)^{\lambda_s/2} Q_0$, where $Q_0=1$~GeV. $\gamma_s$, $\kappa_s$, $\lambda_s$ are parameters to be determined by inclusive DIS data \cite{Chekanov:2001qu,Abramowicz:1900rp,Abramowicz:2015mha}. $\mathcal{A}$ and $\mathcal{B}$ should be evaluated by continuity conditions at $rQ_s = 2$. We use the parametrization
by Soyez \cite{Soyez:2007kg} and two parametrizations in Ref.~\cite{Rezaeian:2013tka} for this investigation, which we provide in
Table~\ref{tab:bCGC}. Note that different conventions were used for the impact parameter dependence in
Refs.~\cite{Soyez:2007kg,Rezaeian:2013tka}. We follow the prescription in Refs.~\cite{Watt:2007nr,Rezaeian:2013tka} for the skewedness
correction in the bCGC dipole model. $R_\text{bCGC}$ is assumed to be,
\begin{eqnarray}
  R_\text{bCGC}(\delta_\text{bCGC}) = \frac{2^{2\delta_\text{bCGC}+3}}{\sqrt{\pi}}
\frac{\Gamma(\delta_\text{bCGC}+5/2)}{\Gamma(\delta_\text{bCGC}+4)} \quad\text{with}\quad \delta_\text{bCGC}  \equiv
\frac{\partial\ln\left(\mathcal{A}_{T,L}^{\gamma^* p\rightarrow Ep}\right)}{\partial\ln(1/x)}.
 \label{eq:Rg_bCGC}
\end{eqnarray}
The obtained $R_\text{bCGC}$ is then applied multiplicatively to the production amplitude.

The parametrizations of dipole model bSat I-III \cite{Kowalski:2006hc} and bCGC I \cite{Soyez:2007kg} were fitted to the 2001 HERA DIS structure function data \cite{Chekanov:2001qu}. The parametrizations of dipole model bSat IV $\&$ V \cite{Rezaeian:2012ji} and bCGC II $\&$ III \cite{Rezaeian:2013tka} were fitted to the 2013 combined DIS data from the ZEUS and H1 collaborations \cite{Abramowicz:1900rp}. The ZEUS and H1 collaborations released updated combined DIS data in 2015 with more data points and higher precision \cite{Abramowicz:2015mha}. A recent study \cite{Ahmady:2016ujw} shows that the CGC dipole model parametrization of Ref.~\cite{Rezaeian:2013tka} gives an excellent fit to the 2015 combined DIS data with $\chi^2/\text{d.o.f.}=1.07$, and a refit to the 2015 combined DIS data gives parameters rather similar to those fitted to the 2013 combined DIS data. Thus we expect only small discrepancies between parametrizations fitted to the 2013 combined DIS data and 2015 combined DIS data. In Sections \ref{sec:hera} and \ref{sec:upc} we show results with the dipole parametrization bCGC II \& III, which were fitted to the 2013 combined DIS data \cite{Abramowicz:1900rp}, and in Section \ref{sec:model} we compare all $8$ parametrizations of Tables \ref{tab:bsat} and \ref{tab:bCGC}.

\subsection{Heavy quarkonium in a holographic basis} 
\label{ssec:BLFQ}

Most phenomenological vector meson LFWFs used in the literature are based on analogy with the virtual photon LFWF, which can be evaluated perturbatively \cite{Lepage:1980fj,Forshaw:2003ki}. Phenomenological
wavefunctions typically have the same spin structure as the photon LFWF, and differ only by the specification of the scalar components of the LFWFs. For example, the boosted Gaussian (bG)
\cite{Brodsky:1980vj,Nemchik:1996cw} LFWFs are obtained by boosting a Gaussian type wavefunction in the meson rest frame to the infinite momentum frame. Our vector meson LFWFs are obtained by solving for the charmonium bound states of an effective Hamiltonian \cite{Li:2015zda}. The dynamics of this effective Hamiltonian determines the spin structure of the bound states; in particular, the one-gluon exchange interaction gives rise to D-wave components in our vector meson LFWFs.

Our effective Hamiltonian is based on the correspondence between anti-de Sitter (AdS) space and QCD, which leads to the light-front holographic QCD \cite{deTeramond:2005su,Brodsky:2014yha}. In the light quark sector the light-front holographic QCD wave functions lead to diffractive $\rho$ and $\phi$ electroproduction that are in agreement with HERA data \cite{Forshaw:2012im, Ahmady:2016ujw}. It is a challenge to apply the light-front holographic QCD to the heavy flavor sector since light-front holographic QCD works only in the zero or small quark mass limit \cite{Brodsky:2014yha}.

Within the basis light-front quantization formalism, Li {\it et al.} generalized the light-front holographic QCD in Ref.~\cite{Li:2015zda} by introducing a longitudinal confining potential and including the one-gluon exchange dynamics. The heavy quarkonia spectroscopy and the corresponding LFWFs are obtained by solving the light-front Schr\"{o}dinger equation with the effective Hamiltonian,
\begin{eqnarray}
\label{eq:Ham}
 H_\text{eff} = \frac{\bm k^2 + m_q^2}{z(1-z)}  + \kappa_\text{con}^4 \bm \zeta^2 
 - \frac{\kappa_\text{con}^4}{4 m_q^2}\partial_z \big(z(1-z) \partial_z \big) 
 -\frac{4\pi C_F \alpha_s}{Q^2} \bar u_{s}(k)\gamma_\mu u_{s'}(k') \bar v_{\bar s'}(\bar k') \gamma^\mu v_{\bar s}(\bar k)\;,
 \end{eqnarray} 
where $C_F=\frac{4}{3}$, $Q^2=-\frac{1}{2}(k'-k)^2-\frac{1}{2}(\bar k-\bar k')^2$. The last term is the one-gluon exchange interaction 
derived from the light-front QCD, and provides the short-distance physics and spin structures needed for the angular excitations and the hyperfine structure. The rest of
the Hamiltonian is developed based on light-front holographic QCD \cite{deTeramond:2005su,Brodsky:2014yha}, which dominates the long-distance
physics and delivers an effective confinement. The longitudinal confining potential in
Eq.~(\ref{eq:Ham}) was proposed for heavy quarkonia in Ref.~\cite{Li:2015zda}, implementing the pQCD
asymptotics for the distribution amplitude (DA) at the endpoints, $\phi^\textsc{da}(x)\sim x^\alpha(1-x)^\beta$. In Ref.~\cite{Li:2015zda}, the model is solved in BLFQ with the (generalized) light-front holographic wavefunctions $\phi_{nm}(\bm k/\sqrt{z(1-z)})$ and $\chi_l(z)$ being adopted as the basis
functions:
\begin{equation}
 \langle \bm k, z, s, \bar s|\psi_h\rangle \equiv \psi_{s\bar s}(\bm k, z) = \sum_{n, m, l} f_{nmls\bar s} \; \phi_{nm}(\bm
k/\sqrt{z(1-z)})\chi_l(z),
\end{equation}
where $\phi_{nm}$ and $\chi_l$ are analytically known functions\footnote{In particular, the soft-wall wavefunction $\phi_{nm}$ is the
harmonic oscillator function in holographic variable $\bm k/\sqrt{z(1-z)}$, which is a generalization of the boosted Gaussian
wavefunction.} and the coefficients $f_{nmls\bar s}$ are obtained through diagonalization.
 
The model for the effective Hamiltonian has several parameters. The strong coupling constant $\alpha_s$ is fixed, $\alpha_s(M_{c\bar c}) \simeq 0.36$ and 
$\alpha_s(M_{b\bar b}) \simeq 0.25$, related via pQCD evolution of the coupling constant. The effective quark mass $m_q$ and the confining
strength $\kappa_\text{con}$ are determined by fitting the mass spectrum of the Hamiltonian to the experimental spectrum for heavy quarkonium states below the open-flavor thresholds. Thus the charmonium spectrum is fitted to $8$ states ($2$ of which are vector mesons), with fit parameters $m_c = 1.522$~GeV and $\kappa_\text{con} = 0.938$~GeV, and bottomonium is fitted to 14 states ($4$ of which are vector mesons) with fit parameters $m_b = 4.763$~GeV and $\kappa_\text{con} = 1.490$~GeV. Both fits have a root-mean-square deviation in their masses from experiment of about $50$~MeV. The
resulting LFWFs are used to calculate the decay constants, the form factors and the charge radii \cite{Li:2015zda}. The results compare
reasonably well with the experiments and other established methods (Lattice QCD and Dyson-Schwinger Equations). Here we use these same LFWFs of the $J/\Psi$ and $\Psi(2s)$ for the calculation of diffractive vector meson production, without adjusting the parameters.

\begin{figure}
 \centering 
\includegraphics[width=.40\textwidth]{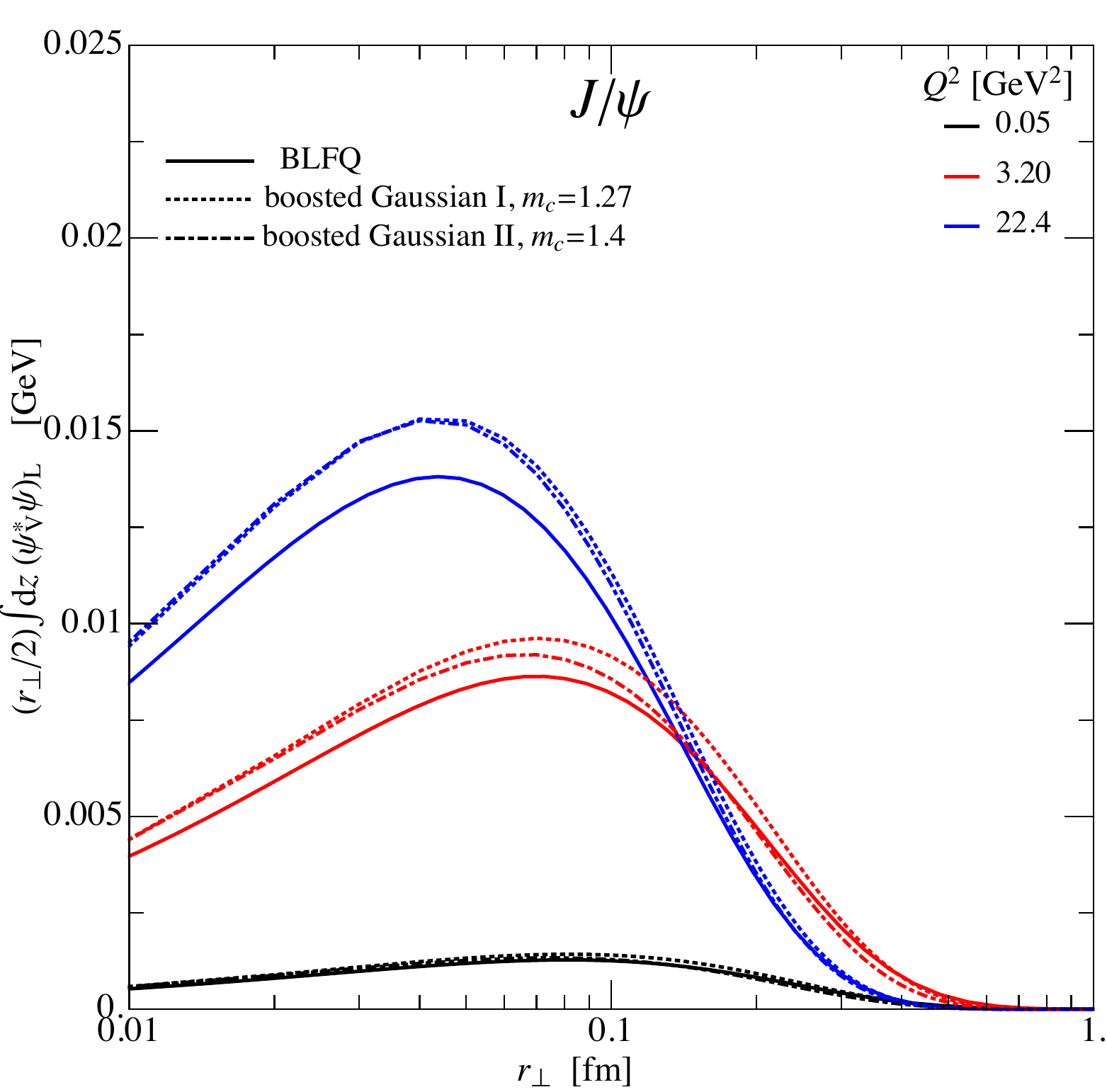}
\includegraphics[width=.40\textwidth]{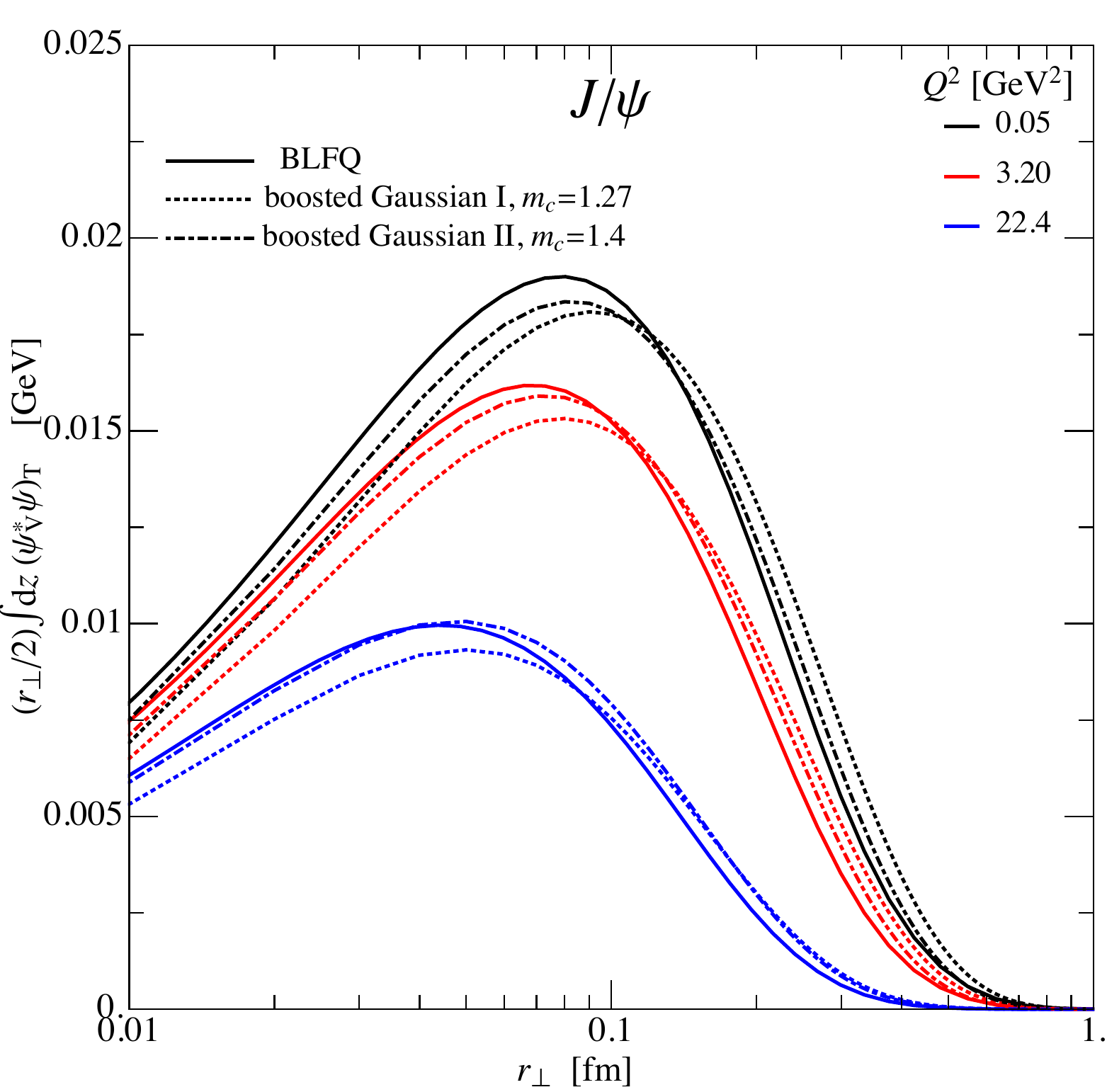}
\caption{(Colors online) The longitudinal (\textit{Left}) and transverse (\textit{Right}) overlap functions between the $J/\Psi$
LFWF and the photon LFWF predicted by the BLFQ LFWF and two parametrizations of the boosted Gaussian LFWF for three representative values of $Q^2$. The BLFQ LFWF is
obtained by diagonalizing the effective Hamiltonian in Eq.~(\ref{eq:Ham}) \cite{Li:2015zda}. The boosted
Gaussian LFWF I and II are given in Refs.~\cite{Armesto:2014sma} and \cite{Kowalski:2006hc}, respectively. The charm quark mass in the virtual photon LFWF is set to $1.27$~GeV when calculating the overlap function between the BLFQ LFWFs and virtual photon LFWF, see texts.
}
\label{fig:overlap}
\end{figure}

We interpret the quark mass obtained by the fitting as the effective quark mass in the bound state, which is not necessarily the same as the quark mass in the virtual photon LFWF or the dipole cross section. In this investigation, we set the quark mass in the virtual photon LFWF to be $1.27$~GeV when calculating the overlap function between the BLFQ LFWFs and virtual photon LFWF. In Fig.~\ref{fig:overlap},  we present the overlap function between the  $J/\Psi$ LFWF and the photon LFWF integrated over $z$ as predicted by the BLFQ LFWF and two parametrizations of the boosted Gaussian LFWF as a function of transverse separation of the quark and antiquark. The BLFQ LFWF is obtained by diagonalizing the effective Hamiltonian in Eq.~(\ref{eq:Ham}), as outlined above \cite{Li:2015zda}. The boosted Gaussian LFWFs with charm quark mass equals $1.27$~GeV (boosted Gaussian I) and $1.4$~GeV (boosted Gaussian II) are given in Ref.~\cite{Armesto:2014sma} and Ref.~\cite{Kowalski:2006hc}, respectively. Parametrizations with different charm quark masses generate significantly different results \cite{Kowalski:2006hc, Rezaeian:2013tka}. In this investigation, we focus on comparing the prediction of the BLFQ LFWF to experiments, and use the predictions of boosted Gaussian model for comparisons.

\begin{figure}
 \centering 
\includegraphics[width=.40\textwidth]{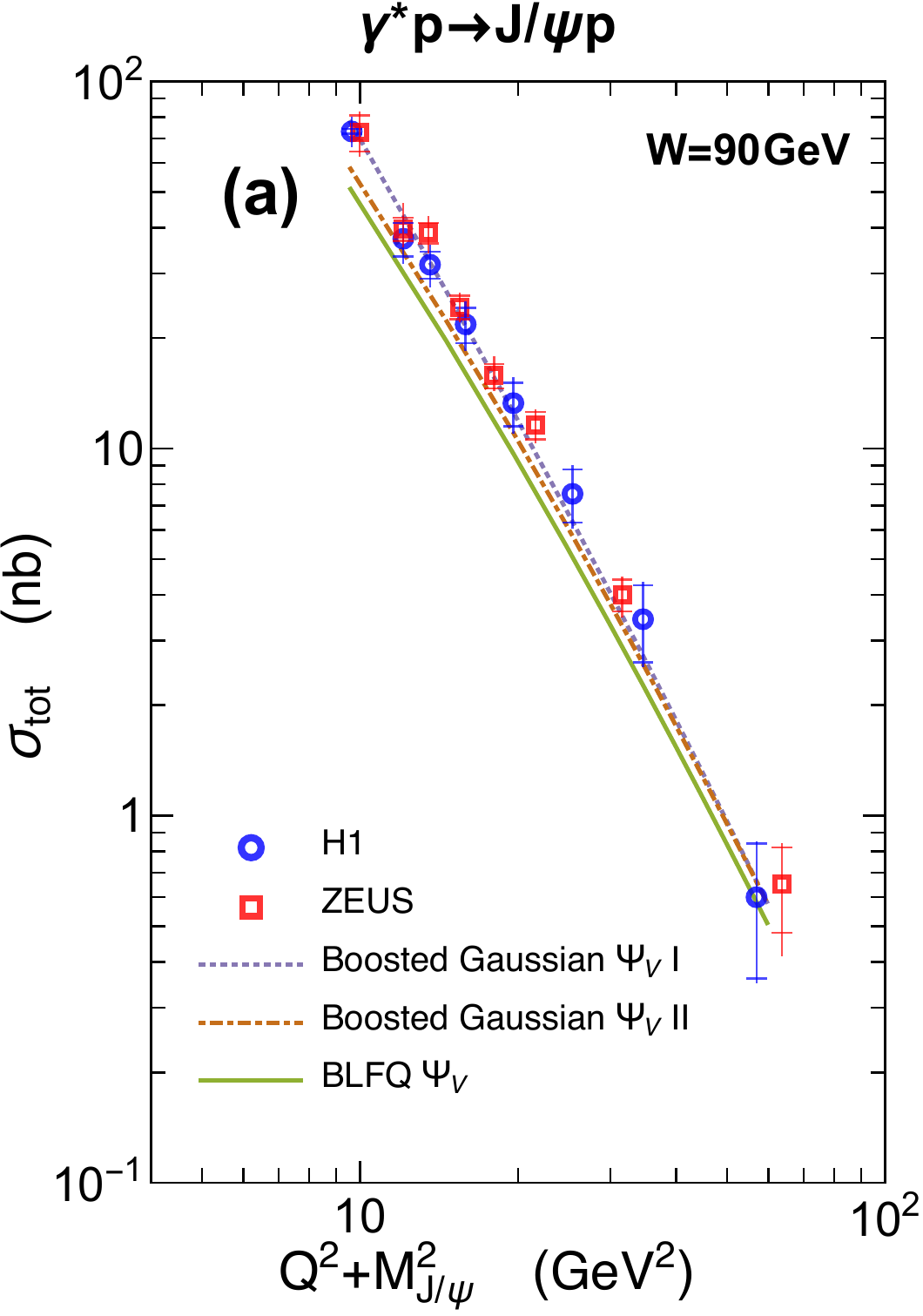}
\raisebox{0.02\height}
{\includegraphics[width=.395\textwidth]{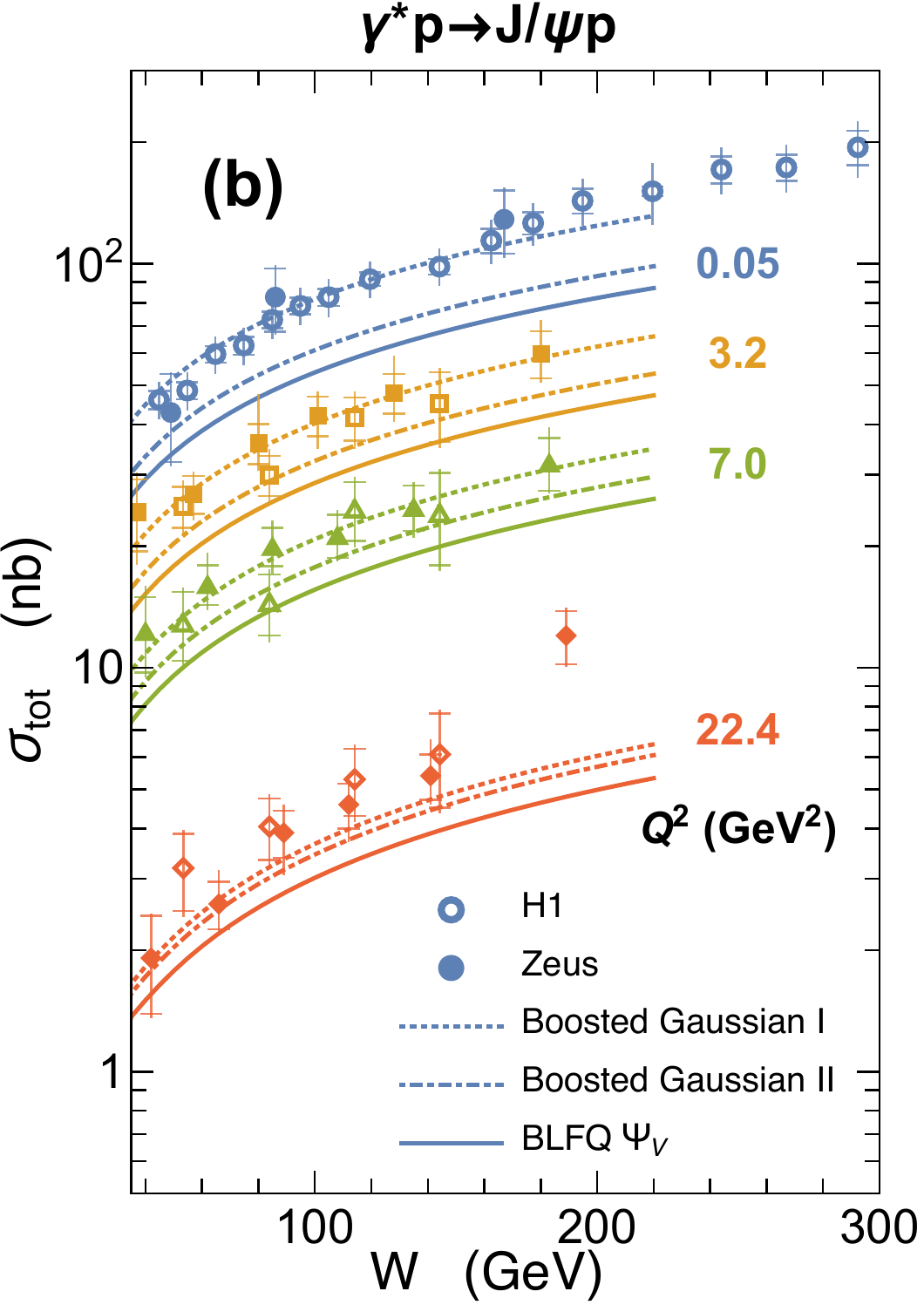}}
\includegraphics[width=.40\textwidth]{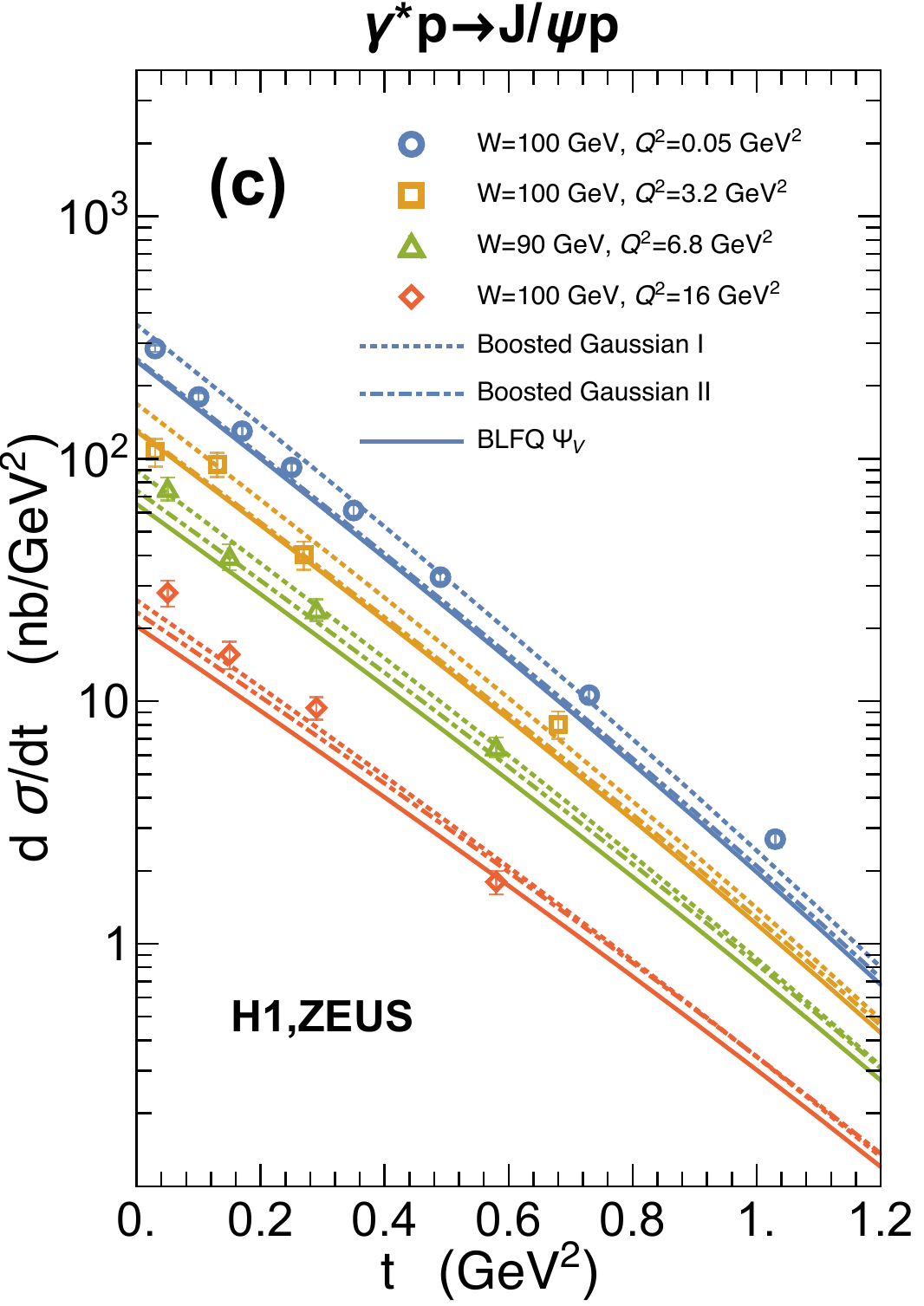}
\includegraphics[width=.385\textwidth]{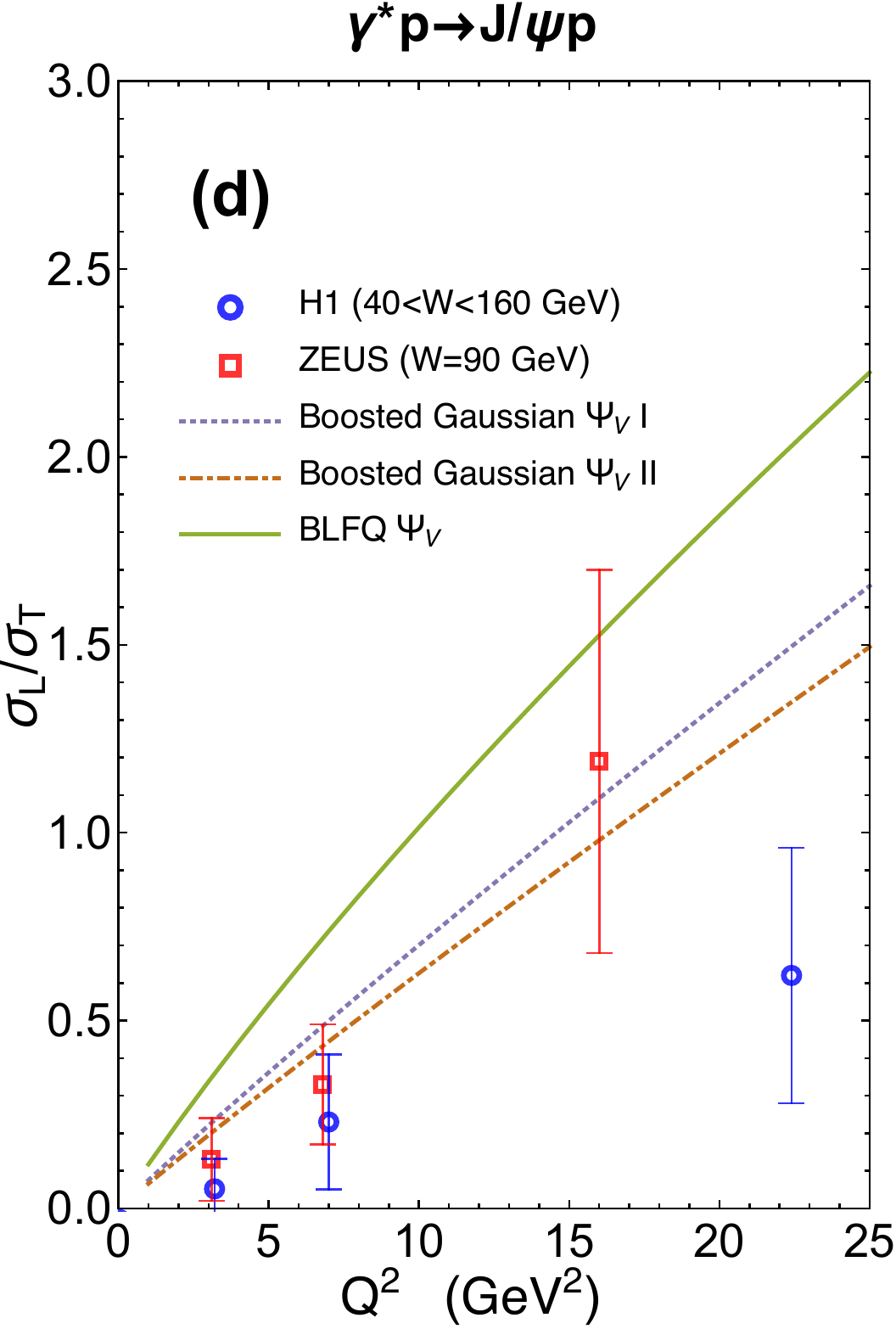}
\caption{(Colors online) Predictions of the BLFQ LFWF (solid curves, using bCGC III in Table~\ref{tab:bCGC}), the boosted Gaussian I LFWF \cite{Armesto:2014sma} (dotted curves, using bCGC II in Table~\ref{tab:bCGC}) and the boosted Gaussian II LFWF \cite{Kowalski:2006hc} (dot-dashed curves, using bCGC III in Table~\ref{tab:bCGC}) compare to the HERA experimental data \cite{Chekanov:2004mw,Aktas:2005xu}. The inner bars indicate the statistical uncertainties; the outer bars are the statistical and systematic uncertainties added in quadrature. (a): Total $J/\Psi$ cross section for different value of $(Q^2+M_V^2)$ at $W=90$~GeV. 
(b): Total $J/\Psi$ cross section for different values of $Q^2$ and $W$. (c): The $J/\Psi$ differential cross section $\dd \sigma/ \dd t$ as a function of $t$. 
(d): The ratio of the longitudinal to the transverse cross sections $R=\sigma_L/\sigma_T$ as a function of $Q^2$.  
}
\label{fig:hera}
\end{figure}

\section{Charmonium production at HERA} 
\label{sec:hera}

We calculate charmonium production using five sets of parameters in the bSat dipole model \cite{Kowalski:2006hc,Rezaeian:2012ji} and three
sets of parameters in the bCGC dipole model \cite{Soyez:2007kg,Rezaeian:2013tka} using the BLFQ charmonium LFWF in the kinematic range of
the HERA experiment \cite{Chekanov:2004mw,Aktas:2005xu,Abramowicz:2016ext}. Various cross sections obtained as a function of the kinematic
variables $Q^2$, $W$ and $t$ are in reasonable agreement with experimental data. As an illustration, we present some representative results in Fig.~\ref{fig:hera}, together with calculations using boosted Gaussian wavefunctions for comparison. In all four panels the solid curves are calculated with our BLFQ vector meson LFWF, the dotted curves are calculated with the boosted Gaussian I LFWF of Ref.~\cite{Armesto:2014sma} with $m_c=1.27$~GeV, and the dot-dashed curves are calculate with the boosted Gaussian II LFWF of Ref.~\cite{Kowalski:2006hc} with $m_c=1.4$~GeV, respectively. The bCGC III parametrization for dipole cross section was used for BLFQ LFWF and boosted Gaussian II LFWF. The bCGC II parametrization for dipole cross section was used for the boosted Gaussian I LFWF.

Fig.~\ref{fig:hera}(a) shows the total $J/\Psi$ cross section as function of $(Q^2+M_V^2)$ for photon-proton c.m. energy $W=90$~GeV. In Fig.~\ref{fig:hera}(b) we show the total $J/\Psi$ cross section as function of $W$ at various values of $Q^2$. The differential cross section $\dd \sigma/\dd t$ is shown in Fig.~\ref{fig:hera}(c) as function of the momentum transfer $t$. Qualitatively, both the boosted Gaussian LFWFs and the BLFQ LFWF provide reasonable descriptions to the $J/\Psi$ cross section data at HERA. (Note that the boosted Gaussian II LFWF parametrization gives quantitatively better fits to the $J/\Psi$ cross section measurements at HERA, if the bSat I parametrization is used for the dipole cross section \cite{Kowalski:2006hc}.) The BLFQ LFWF calculation underestimates the $J/\Psi$ production at HERA, especially in the small $Q^2$ regime, and the boosted Gaussian LFWFs lead generally to better agreement with the total cross section data. The sizable discrepancy with the HERA measurements at small $Q^2$ should not pose a major hindrance for the application of the BLFQ LFWF to 
diffractive charmonium production. The theoretical uncertainty in the dipole model is large at small $Q^2$. For the dipole cross
section we employed, the photon wavefunction is calculated based on tree-level perturbative QED, without taking confinement and QCD
corrections into account. The pQED photon wavefunction is more reliable at large spacelike values of $Q^2$, since contributions from large
size dipoles are suppressed except at the end points of $z$. At small $Q^2$, confinement is likely to play an important role
\cite{Forshaw:2003ki}. For these reasons, the $J/\Psi$ cross section at small $Q^2$ may have a stronger model dependence. Such uncertainty may be
reduced by a consistent treatment of the heavy quarkonium wavefunction and the photon wavefuntion, e.~g., by including confinement and QCD corrections in the photon wavefunction.

Finally, in Fig.~\ref{fig:hera}(d) we show the ratio of the longitudinal to transverse cross section, $R=\sigma_L/\sigma_T$, for $J/\Psi$ production at HERA. We find a qualitative difference between the boosted Gaussian LFWFs and our BLFQ LFWF. The current data at large $Q^2$ seem to favor the boosted Gaussian LFWF, but the error bars are large.

\begin{figure}
 \centering 
\includegraphics[width=.45\textwidth]{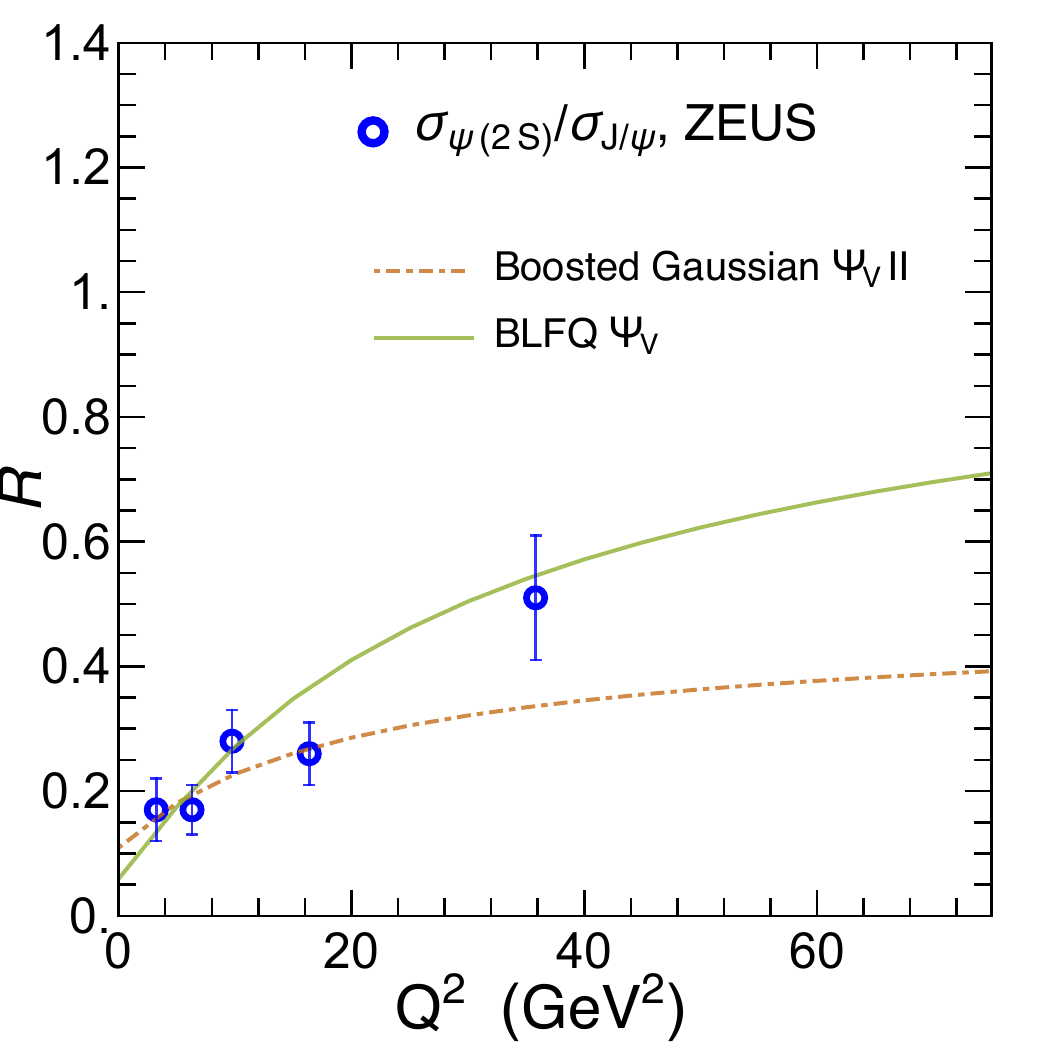}
\caption{(Colors online) Using the bCGC III parametrization in Table~\ref{tab:bCGC}, the predictions of the BLFQ LFWF (solid curve) and the boosted Gaussian II LFWF \cite{Kowalski:2006hc} (dot-dashed curve) for the cross-section ratio $\sigma_{\Psi(2s)}/\sigma_{J/\Psi}$ as a function of $Q^2$ compare to the HERA experimental data \cite{Abramowicz:2016ext}.
}
\label{fig:ratio}
\end{figure}

In Fig.~\ref{fig:ratio}, we compare the predictions of the BLFQ LFWF (solid curve) and the boosted Gaussian II LFWF (dot-dashed curve) for the cross-section ratio $\sigma_{\Psi(2s)}/\sigma_{J/\Psi}$  as a function of $Q^2$ in electron-proton scattering measured at HERA \cite{Abramowicz:2016ext}. The parameterizations of the boosted Gaussian II can be found in Refs.~\cite{Kowalski:2006hc, Lappi:2014eia}. In the boosted Gaussian model, the parametrization of $\Psi(2s)$ LFWFs are very sensitive to the parametrization of $J/\Psi$ LFWFs. The growing differences between the BLFQ and the boosted Gaussian results at larger $Q^2$ suggest the importance of additional data in this region to further distinguish between the LFWFs. 

An accurate dipole cross section requires quantitative understanding of saturation in the small-$x$ regime, which is not available
currently. Observables with weak dipole cross section dependence are favored if one wants to test the validity of heavy quarkonium LFWFs. We
will discuss the dipole model dependence of the charmonium cross section and cross-section ratio in Sec.~\ref{sec:model}.   

\section{Charmonium production at RHIC and LHC} 
\label{sec:upc}

Diffractive charmonium production processes also occur in the ultra-peripheral heavy-ion collisions (UPC), in which two heavy ions scatter at
large impact parameters. In the rest frame of one of the ions, the exclusive heavy quarkonium production in UPC can be regarded as a result
of the scattering of equivalent photons radiated by the incident ion off the target ion \cite{Bertulani:2005ru,Baltz:2007kq}. At low $x$ the color dipole scatters
coherently from the entire nucleus. The saturation effect is amplified by the large number of nucleons interacting with the dipole along its
path through the nucleus. 

The coherent diffractive heavy quarkonium production can be obtained by averaging over all possible states of nucleon configurations 
$\Omega$ of the nucleus. In the photon-nucleus collision, the corresponding diffractive heavy quarkonium production amplitude can be
calculated by replacing the photon-proton dipole cross section $\dif\sigma_{q\bar q}/\dif^2\bm{b}$ in Eq.~(\ref{eq:newampvecm}) with the
nucleon configuration averaged dipole cross section \cite{Kowalski:2003hm},
\begin{eqnarray}
  \left<\frac{{\rm d}\sigma_{q\bar q}}{{\rm d}^2{\bm{b}}}\right>_\Omega = 
  2\left[1-\left(1-\frac{T_A(\bm{b})}{2}\sigma^p_{q\bar q}\right)^A\right] \; ,
  \label{eq:analytical}
\end{eqnarray}
where $\sigma^p_{q\bar q}$ is the photon-proton dipole cross section, integrated over the impact parameter, and $A$ is the atomic number of
the nucleus. $T_A$ is the thickness profile of the nucleus in the transverse dimension, with $\bm b$ the impact parameter of the dipole
relative to the nucleus. We adopt a Woods-Saxon profile for the nucleon distribution. 

Extensive experimental data have been collected for heavy quarkonium production at both RHIC and LHC. The dipole picture has provided a
reasonable explanation for the experimental data using phenomenological charmonium LFWFs, e.g., Refs.~\cite{Kopeliovich:2001ee,Lappi:2013am,
Ducati:2013bya, Goncalves:2014wna}. In this section, we investigate the predictions of the BLFQ LFWF in comparison with measurements in
Refs.~\cite{Afanasiev:2009hy,Takahara:2013kja,Abbas:2013oua,Adam:2015sia,Khachatryan:2016qhq} using bCGC III
in Table~\ref{tab:bCGC} as dipole cross section parametrization. We also present the prediction of the boosted Gaussian LFWF for comparison.

The latest measurement at RHIC for coherent $J/\Psi$ production at mid-rapidity with two gold nuclei colliding at 
$\sqrt{s_{\text{NN}}}=200$~GeV provides a $J/\Psi$ cross section $\dd \sigma/\dd y = 45.6 \pm 13.3$ (stat) $\pm 5.9$ (sys) $\mu$b
\cite{Takahara:2013kja}. A previous study using the boosted Gaussian II LFWF \cite{Kowalski:2006hc} and bSat I dipole parametrization provides a prediction of $109$
$\mu$b \cite{Lappi:2013am}. Using the BLFQ LFWF, the bCGC III
dipole parametrization provides a prediction of $60.4$~$\mu$b, which is consistent with the latest data within
experimental uncertainty. Note that the  kinematic region of the RHIC experiment corresponds to a photon-nucleon c.m.\ energy of $W=34$~GeV,
with the probed gluons carry $x$ roughly $0.015$, which implies that the dipole picture is marginally applicable for such a process at mid-rapidity.   

\begin{figure}
 \centering 
\includegraphics[width=.45\textwidth]{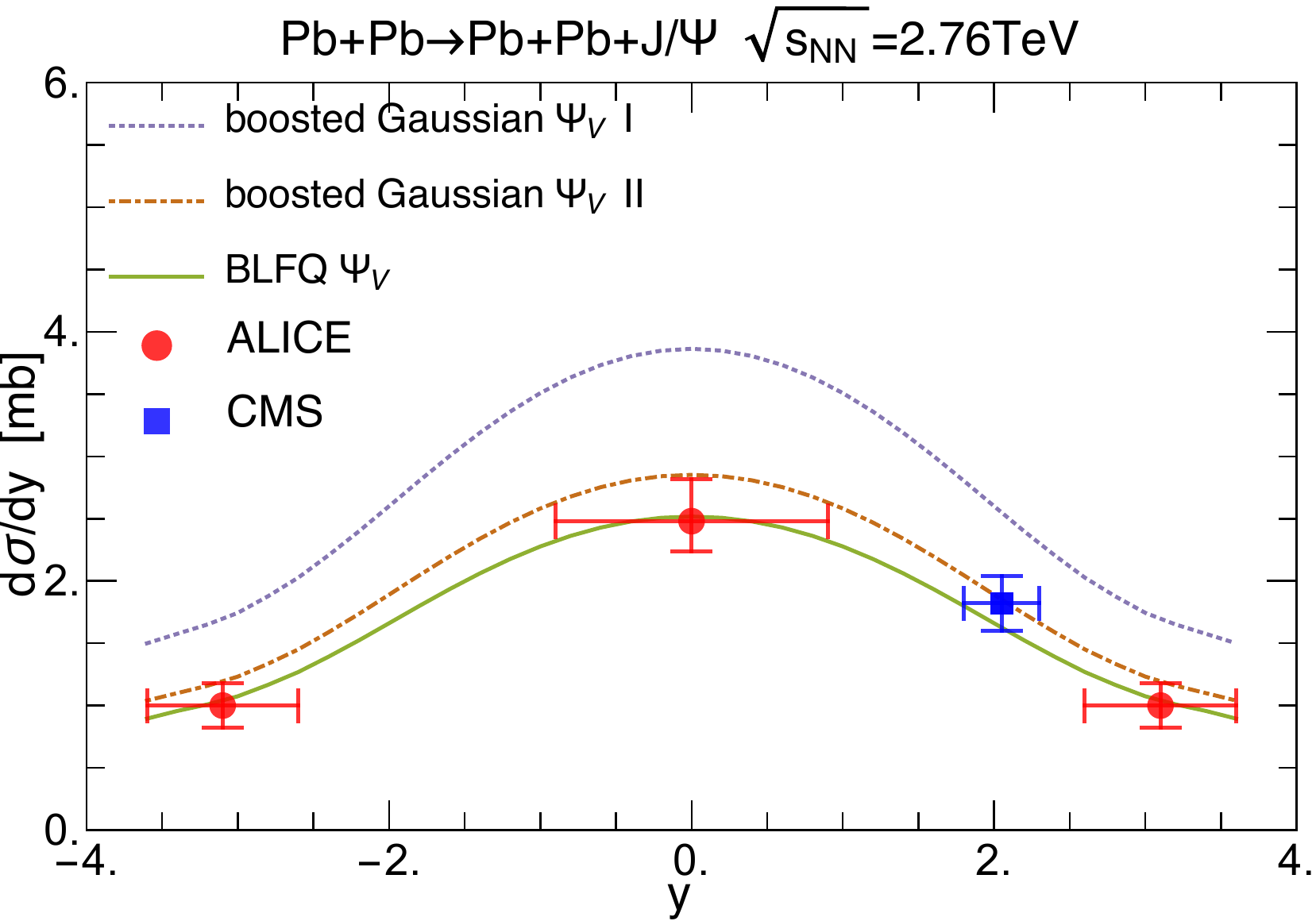}
\caption{(Colors online) The predictions of the BLFQ LFWF (solid curve, using bCGC III in Table~\ref{tab:bCGC}), the boosted Gaussian I LFWF \cite{Armesto:2014sma} (dotted curve, using bCGC II in Table~\ref{tab:bCGC}) and the boosted Gaussian II LFWF \cite{Kowalski:2006hc} (dot-dashed curve, using bCGC III in Table~\ref{tab:bCGC}) for the coherent production of $J/\Psi$ production in Pb-Pb ultra-peripheral collision at $\sqrt{s_{NN}}=2.76$~TeV, compare with the measurements by the ALICE collaboration \cite{Abbas:2013oua, Adam:2015sia} and CMS collaboration \cite{Khachatryan:2016qhq} at LHC. Error bars show statistical uncertainties only.
}
\label{fig:lhc}
\end{figure}

The predictions of the BLFQ LFWF are also consistent with experimental data for coherent production of $J/\Psi$ at mid-rapidity in Fig.~\ref{fig:lhc}. Here, the solid curve, the dotted curve and the dot-dashed curve show the predictions of the BLFQ LFWF, the boosted Gaussian I LFWF \cite{Armesto:2014sma} and the boosted Gaussian II LFWF \cite{Kowalski:2006hc} respectively, for the coherent production of $J/\Psi$ production in Pb-Pb ultra-peripheral collision at $\sqrt{s_{NN}}=2.76$~TeV, compared to the measurements of the ALICE collaboration \cite{Abbas:2013oua, Adam:2015sia} and the CMS collaboration \cite{Khachatryan:2016qhq} at LHC. The bCGC III dipole model
parametrization is implemented for the BLFQ and the boosted Gaussian II LFWFs calculations, the bCGC II dipole model
parametrization is implemented for the boosted Gaussian I LFWF calculation. The predictions of the BLFQ LFWF and the boosted Gaussian II LFWF are within the statistical uncertainty of the experimental data. The prediction of the boosted Gaussian I LFWF slightly overshoots the data.

The cross-section ratio $\sigma_{\Psi(2s)}/\sigma_{J/\Psi}$ measured by the ALICE experiment is approximately twice as large as in photon-proton collision experiments \cite{Adam:2015sia}. The predictions of the BLFQ LFWF, the boosted Gaussian I LFWF and the boosted Gaussian II LFWF, which are based on photon-proton collisions, underestimate $\Psi(2s)$ production in Pb-Pb ultra-peripheral collision at $\sqrt{s_{NN}}=2.76$~TeV. It is possible that $\Psi(2s)$ production is enhanced in photon-nucleus collisions due to nuclear effects. However, more experiments will be needed to confirm this\footnote{Recently, the ALICE collaboration presented their preliminary analysis on $\Psi(2s)$ production for Pb-Pb collision at LHC based on data collected from Run II. The updated cross section ratio $\sigma_{\Psi(2s)}/\sigma_{J/\Psi}$ is consistent with the HERA measurement.}.

\section{Dipole model dependence} 
\label{sec:model}
In the dipole model, the cross section for the diffractive vector meson production is calculated through a convolution of the dipole cross
section and the overlap of the vector meson LFWF and the photon LFWF. Consequently, it is expected that uncertainties from both the dipole
cross section and the vector meson LFWF contribute to the uncertainty of the result. Inclusive data from the DIS experiment at HERA used to
determine the dipole cross sections have large uncertainties in the small-$x$ regime, which leads to large uncertainties in the dipole model
parameters. State-of-the-art fits to the heavy quarkonium cross section measurement are not sufficient for a conclusive statement
on either the dipole cross section parametrization or the heavy quarkonium LFWF. For instance, using the boosted Gaussian parametrization
given in Ref.~\cite{Kowalski:2006hc}, the $J/\Psi$ cross section at $Q^2=0$ could differ by as much as $30\%$ using dipole cross section
parametrizations in Table~\ref{tab:bsat}.  

\begin{figure}
 \centering 
 \includegraphics[width=.32\textwidth]{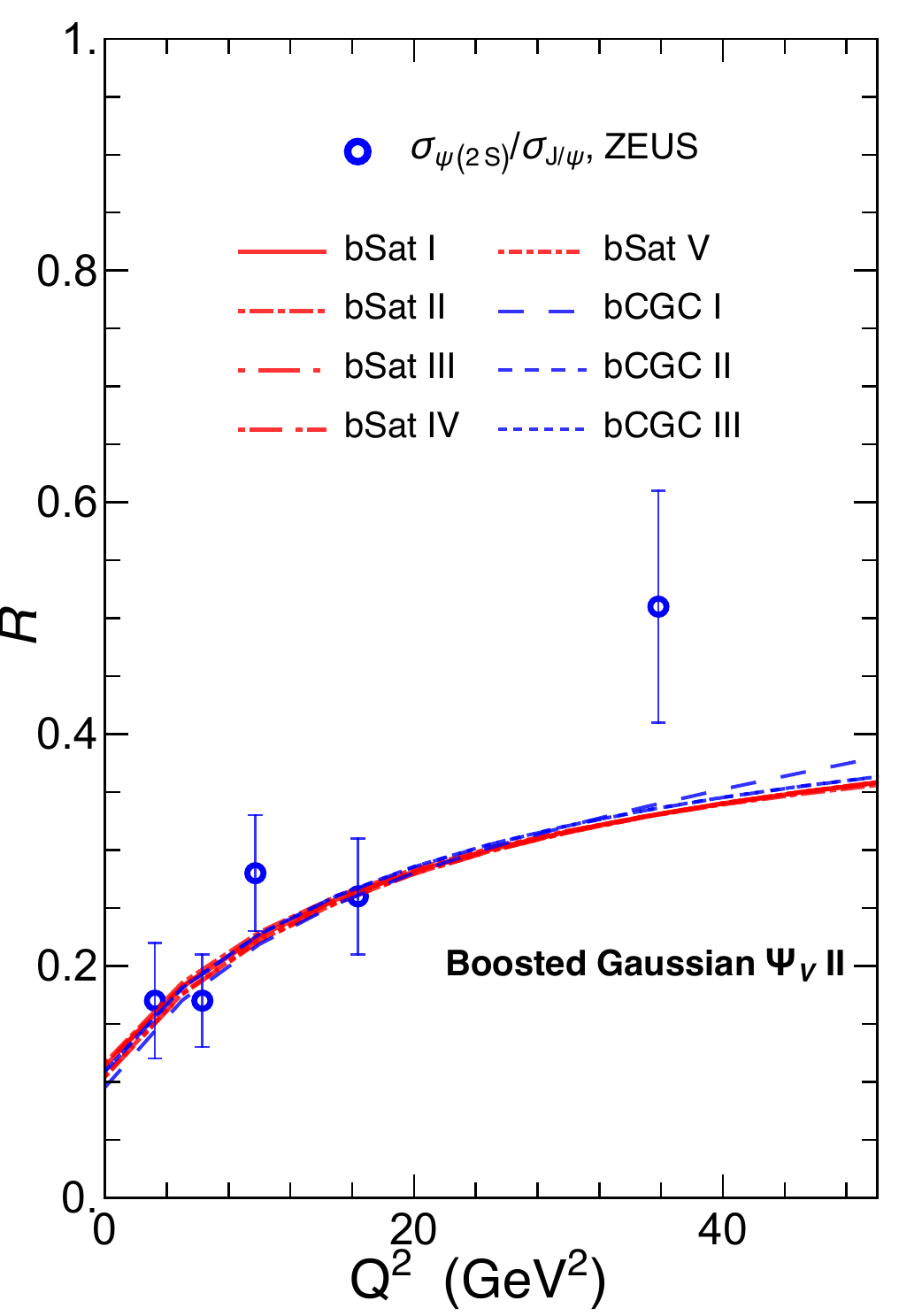}
\includegraphics[width=.32\textwidth]{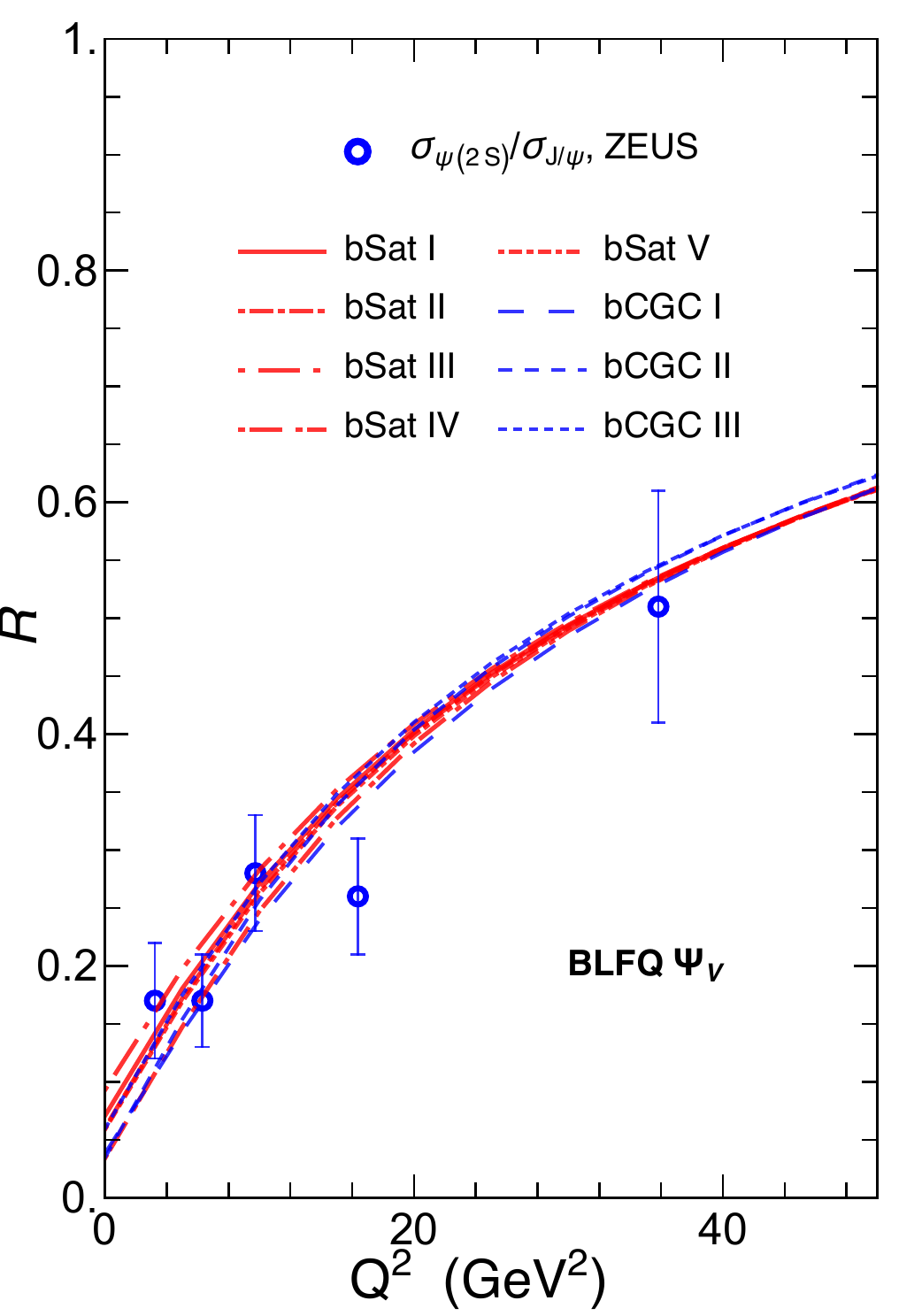}
\includegraphics[width=.32\textwidth]{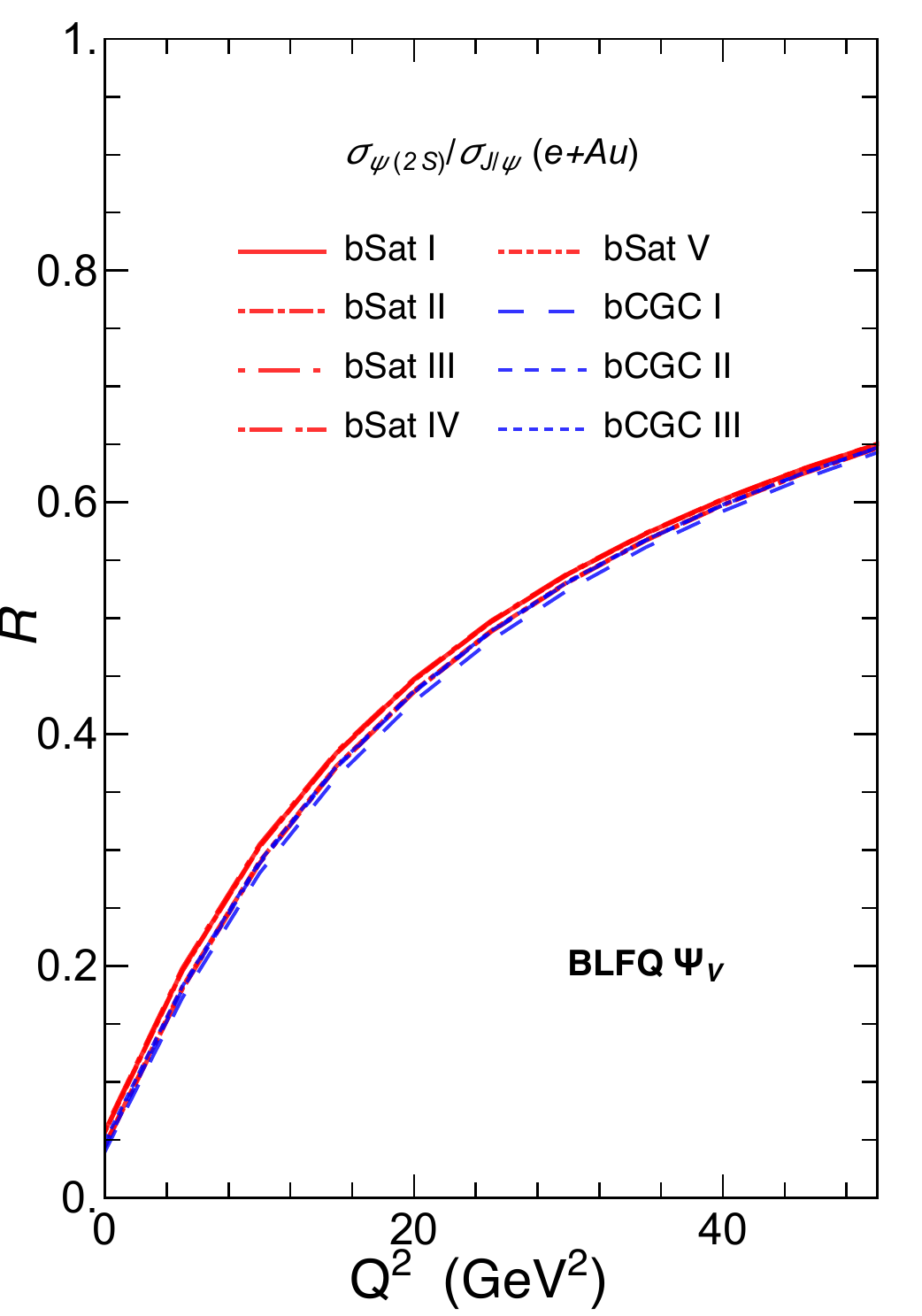}
\caption{(Colors online) The cross-section ratio $\sigma_{\Psi(2s)}/\sigma_{J/\Psi}$ as a function of $Q^2$ predicted by the boosted Gaussian II LFWF \cite{Kowalski:2006hc, Lappi:2014eia} (\textit{Left}) and the BLFQ LFWF (\textit{Middle})
using various dipole cross section parametrizations in Table~\ref{tab:bsat} and \ref{tab:bCGC} for electron-proton collisions. The cross-section ratio $\sigma_{\Psi(2s)}/\sigma_{J/\Psi}$ as a function of $Q^2$ predicted by BLFQ LFWF in the coherent charmonium production in electron-gold collisions is shown in the right panel (\textit{Right}). The cross-section ratio exhibits weak dependence on dipole models. The experimental data points are measurements by the ZEUS collaboration \cite{Abramowicz:2016ext} in electron-proton scattering at HERA.}
\label{fig:model}
\end{figure}    

On the other hand, one might expect that the quark-antiquark pair originating from quantum fluctuation of the virtual photon scatter
universally on the nuclear target for the production of different states of the same quarkonium system, e.g., $J/\Psi$ and $\Psi(2s)$. Under such an assumption, the cross-section ratio of higher excited states over the ground state should exhibit weaker
dependence on the dipole model than the cross section itself. Our calculations suggest this is indeed the case. In Fig.~\ref{fig:model}, we calculate the ratio of the $\Psi(2s)$ cross section to the $J/\Psi$ cross section as a function of $Q^2$ predicted
by the boosted Gaussian II LFWF \cite{Kowalski:2006hc, Lappi:2014eia} (\textit{Left}) and the BLFQ LFWF (\textit{Middle}) using various dipole cross section parametrizations in Table~\ref{tab:bsat} and ~\ref{tab:bCGC} for electron-proton
collisions. The kinematic variables are chosen to be the mean values of experimental measurements \cite{Abramowicz:2016ext}. We observe that
the cross-section ratio exhibits weak dependence on dipole models, especially in the large $Q^2$ regime, where the dipole models we adopted are well motivated by $k_t$ factorization \cite{Gribov:1984tu, Collins:1991ty}. 

In the right panel of Fig.~\ref{fig:model}, we also present the BLFQ LFWF predictions for the cross-section ratio $\sigma_{\Psi(2s)}/\sigma_{J/\Psi}$ in electron-gold collisions, with $\Psi(2s)$ and $J/\Psi$ being produced coherently. We use the
mean value of $W$ at the HERA experiment of $ep$ collisions after integrating over $t$. The cross-section ratio for electron-gold collisions also shows weak dependence on the dipole model. The calculated results change by less than $1\%$ when the gold target is replaced by lead target. We also observe that the cross-section ratio $\sigma_{\Psi(2s)}/\sigma_{J/\Psi}$ predicted by the BLFQ LFWFs is insensitive to the charm quark mass in the virtual photon LFWF.

Future electron-ion collision experiments with high luminosity are expected to deliver data on production of higher excited states of heavy quarkonium over a wide kinematic range \cite{Accardi:2012qut}. The uncertainties associated with heavy quarkonium LFWFs could be reduced through measurements of cross-section ratios of higher excited states to the ground state, owing to the insensitivity of such ratios to the dipole model. With well-constrained heavy quarkonium LFWFs, the gluon density distribution in the small-$x$ regime could be extracted efficiently through the diffractive heavy quarkonium production process.

\section{Summary and Outlook}
\label{sec:sum}

Using established dipole models, we study diffractive charmonium production with a theoretical LFWF obtained from the basis 
light-front quantization approach. One-gluon exchange dynamics from light-front QCD and an effective confining potential inspired by
light-front holographic QCD are implemented in the effective Hamiltonian. Two parameters in the effective Hamiltonian are fixed by the
mass spectrum of charmonium. The resulting charmonium LFWF gives reasonably good descriptions of currently available experimental data at
HERA, RHIC and LHC within the dipole model. We observe that the cross-section ratio of $\sigma_{\Psi(2s)}/\sigma_{J/\Psi}$ as a function of
$Q^2$ has a weak dependence on the dipole model but is rather sensitive to the charmonium wavefunction. We suggest that future electron-ion
collision experiments could reduce theoretical uncertainties associated with the structure of heavy quarkonium by measuring the
cross-section ratios of the higher excited states to the ground state, owing to their weak dependence on the dipole cross section. Accurate
meson wavefunctions will eventually lead to a more precise description of the gluon distribution in the saturation regime 
\cite{Accardi:2012qut}. 

We are currently extending our calculations to diffractive bottomonium production. Note that the masses of the $J/\Psi$ and $\Psi(2s)$ are the only two {\it vector} charmonium states used in the fit of the parameters in the effective Hamiltonian in the BLFQ approach. On the other hand, four {\it vector} bottomonium states, $\Upsilon(1s)$, $\Upsilon(2s)$, $\Upsilon(3s)$ and $\Upsilon(1d)$, were used in the fit of the effective Hamiltonian parameters. Having four BLFQ LFWFs would therefore provide more cross-section ratio predictions in the bottomonium sector and, in principle, lead to improved confidence in extracting  gluon saturation properties.

The effective Hamiltonian, whose wavefunctions we employ, has been fitted only to the mass spectra of heavy quarkonia. The resulting LFWFs 
have been found to provide reasonable descriptions of heavy quarkonia decay constants, form factors and now diffractive meson production.
Therefore, the BLFQ formalism provides a platform for a unified description of the physical observables mentioned above. Our future goals
include improving the effective Hamiltonian of heavy quarkonium using additional experimental measurements, such as decay constants, as constraints. Using the BLFQ
LFWFs to predict diffractive heavy quarkonium production in future electron-ion collision experiments is an important future goal since
a large amount of experimental data, especially data on the production of higher excited heavy quarkonium states, are anticipated.

\section*{Acknowledgments} 
We wish to thank X.~Zhao, P.~Wiecki, A.~Rezaeian and Y.~Xie for valuable discussions and communications. 
We thank N. Kovalchuk for providing us the experimental data for the $\Psi(2s)$ measurement.
This work was supported in part by the Department of Energy under
Grant Nos. DE-FG02-87ER40371 and DESC0008485 (SciDAC-3/NUCLEI). We acknowledge computational resources provided by the National Energy 
Research Scientific Computing Center (NERSC), which is supported by the Office of Science of the U.S. Department of Energy under Contract
No. DE-AC02-05CH11231. 

%


\end{document}